\documentclass[11pt]{article}
\usepackage{graphicx}
\usepackage{hyperref}

\newcommand{\artanh}{\rm \, artanh\,}

\begin{document}
\renewcommand{\abstractname}{}

\baselineskip=20pt         

\thispagestyle{plain}
\begin{center}
\LARGE \bf Is it possible to see the infinite future of\\[7pt]
the Universe when falling into a black hole?
\end{center}

\vspace{4pt}
\begin{center}
\large A.\,A. Grib\footnote{
{\bf A.\,A. Grib} \ A.I.Herzen Russian State Pedagogical University,\\
nab. r. Moiki 48, 191186  St.\,Petersburg, Russian Federation\\
E-mail: \, {andrei\_grib@mail.ru}}, \
Yu.\,V. Pavlov\,\footnote{
{\bf Yu.\,V. Pavlov} \ Institute of Problems in
Mechanical Engineering, Russian Academy of Sciences,\\
V.O., Bol'shoi prosp. 61, 199178  St.\,Petersburg, Russian Federation\\
E-mail: \, yuri.pavlov@mail.ru }
\end{center}

\vspace{-17pt}
    \begin{abstract} {\noindent
\underline{\bf Abstract.} \,
A possibility to see the infinite future of the Universe
by an astronaut falling into a black hole is discussed and ruled out.
}\\[4pt]
 {\small \bf {Methodological Notes.}}\\[7pt]
\end{abstract}


\vspace{7pt}
\section{Introduction}
    Black holes are considered to be quite usual objects in modern
astrophysics.
    There is convincing observational evidence for their existence
(see, for example, review~\cite{Cherepashchuk03}).
    According to the common point of view, there is a black hole at the
galactic center, and black holes reside in quasars and cause their bright
emission due to the `eating' of infalling stars and interstellar gas.
    In contrast to supermassive black holes in galactic nuclei and quasars
with a mass of millions of times the Sun, there are less massive black holes
which are observed in binary systems due to their interaction with
the companion star.
    However, when attempting a theoretical description of a black hole in
the context of General Relativity, some disagreements appear, both in special
and popular literature.
   Because of this, the aim of our notes consists in examining some of these
discrepancies.

    Here we shall follow Einstein's general relativity.
    In alternative theories, for example, in the field theory of
gravitation~\cite{LogunovMst}, there can be no black holes at all.

    There can be static, rotating, and charged black holes.
    They are described by the Schwarzschild (1916)~\cite{Schwarzschild16},
Kerr~\cite{Kerr63}, Reissner--Nordstr\"{o}m~\cite{Reissner16}
(charged nonrotating), and Kerr--Newman~\cite{Newmanetal65}
(charged rotating) metrics, respectively.
    Yet the common point of view is that the charge of a black hole can be
neglected if it was produced from the core collapse of a star consisting of
ordinary nucleons and electrons~\cite{NovikovFrolov}.

    Consider the best studied case of a static black hole.
    What would an astronaut falling into such a black hole see?
    In all textbooks in which general relativity is considered
(see, for example, Ref.~\cite{LL_II}), one can read that there are two frames
of reference.
    The first frame (call it~$\mathbf{A}$) is related to the Earth;
the second one~($\mathbf{B}$) is related to the astronaut falling
upon the black hole.
    In the first frame of reference, the astronaut will forever approach
the surface of black hole (the horizon at the Schwarzschild radius of the
black hole) but never reach it.
    In the second frame, the astronaut will reach the Schwarzschild radius
in a finite time interval and cross the black hole horizon, but any signal
produced by him can never reach an observer on the Earth.
    And here such a non-naive physicist as Yuval Ne'eman asks a naive
question: ``How can~$\mathbf{B}$ be allowed his (or her) frame of
reference, in the equalitarian regime of covariance, if we can claim in all
finality that~$\mathbf{B}$ will never cross the Schwarzschild radius, in our
spacetime reality?''~\cite{Neeman}.
    Similarly, the collapsing star will never cross its Schwarzschild radius
in frame~$\mathbf{A}$.
    Next, Ne'eman asks how one can ``add to eternity~$\mathbf{A}$ an extra
half-hour~$\mathbf{B}$ spends inside the black hole.''
    He calls the emerging situation `surrealism'-- hypothesis for
the existence of different realities, one of which is not only unavailable but
also impossible for another.
    Let us attribute this observation to the problem of the correct `philology'
and accept that the brave astronaut is capable of passing from one reality to
another.

    The situation discussed by Ne'eman is usually described in terms of
the complete and incomplete frames of references.
    For example, frame~$\mathbf{A}$ is incomplete since one cannot describe
there events inside the black hole, while frame~$\mathbf{B}$ in
the Kruskal--Szekeres coordinates~\cite{Kruskal60} is complete.
    This answer, of course, was known to Ne'eman, but apparently was not
fully satisfactory to him.
    The time of the astronaut inside a black hole is in no way related to our
time on the Earth, and, as was mentioned above, it can in no way be `added'
to it.
    Furthermore, there is a purely mathematical problem related to
the singularity of the very transform of passage from~$\mathbf{A}$
to~$\mathbf{B}$ at the Schwarzschild radius, which has been discussed
by theoreticians ever since the appearance of the Schwarzschild black
hole solution~\cite{Eizenshtedt82}.

    Let us agree, however, with the commonly accepted opinion with regard
to the astronaut's  crossing the Schwarzschild radius.
    Let us ask: What will he see when approaching a black hole?
    In the popular literature~\cite{CherCHDV}
(see also Ref.~\cite{AstronomiyaVekXXI}), a very attractive picture for
future tourists to the galactic nucleus is suggested:
the astronaut can see all the future of the Universe.
    ``A spacecraft with astronauts approaching a black hole will appear
to the Earth's observer as breaking its motion but never crossing the black
hole horizon.
    If the situation is reversed and we analyze it from the point of view of
the astronaut lingering near the horizon then the rate of events in the
external Universe is extremely accelerated: virtually in one moment of his
time the astronaut will see the infinitely long development of events
in the external Universe.
    He will see how our Sun expands to become a red giant, how the Earth
evaporated from the hot solar rays when sliding over upper layers of dying
Sun's atmosphere, how the outer hydrogen envelop detaches from the Sun
that ultimately turns into a white dwarf --- in short, the astronaut will
see the future of our Universe!''
    The astronaut will observe all that over a finite time interval
in the frame of reference~$\mathbf{B}$.
    Is that the case?

    A similar statement can be read in the translator's notes to
book~\cite{HawkingUnNut}, ``explaining'' to the reader considerations
of the author, Stephen Hawking (!).
    The same picture for an observer sitting on the surface of a collapsing
star is suggested in the popular book~\cite{Redze}:
``It appears to such an observer that the time in the external space runs
at a growing rate and instantly reaches the very `end of all times'.''
    Unfortunately, we must disappoint future astronauts and popular book
readers.
    The astronaut falling upon a black hole is never seeing the infinite
future of our Universe!
    To clarify this, let us write out several formulas.

\vspace{11pt}
\section{Free fall upon a Schwarzschild black hole}

    Consider free fall upon a static noncharged black hole in
Schwarzschild coordinates in which the metric has the form
    \begin{equation} \label{Sch}
ds^2=\left(1-\frac{r_{\! \rm g}}{r} \right) c^2 dt^2 - \frac{dr^2}{
 \displaystyle 1- r_{\! \rm g}/r } -
r^2 \left( d \vartheta^2 +\sin^2 \vartheta \, d \varphi^2 \right).
\end{equation}
Here, $r_{\! \rm g}= 2 G m / c^2 $ is the gravitational radius of the black hole,
and\, $ c $\, is the speed of light.

    Radial geodesics in metric~(\ref{Sch}) satisfy the equations
(see Ref.~\cite{Chandrasekhar})
    \begin{equation} \label{SG}
\biggl( \frac{d r}{c\, d \tau} \biggr)^2 = \frac{r_{\! \rm g}}{r} +
\varepsilon^2 -1 \,, \ \ \ \
\frac{d t}{d \tau} = \frac{\varepsilon}{1 - r_{\! \rm g}/r} \,,
\end{equation}
where $\varepsilon={\rm const}$.
    For timelike geodesics, $\tau$ is the proper time of a moving particle,
and $\varepsilon $ is the specific energy:
a particle with the rest mass $m_0$ possesses total energy
$\varepsilon m_0 c^2 $ in the gravitational field~(\ref{Sch}).

    If the particle's fall starts from rest at some distance
$r_0 > r_{\! \rm g}$ then clearly
(see the first formula in Eqs.~(\ref{SG}) at $dr/d \tau =0$)
\ $\varepsilon=\sqrt{1- r_{\! \rm g}/r_0}$ \ and, hence (after dividing
the first equation by the square of the second one and extracting the root)
    \begin{equation} \label{drdt}
\frac{dr}{c\, dt} = - \left( 1- \frac{r_{\! \rm g}}{r} \right)
\Biggl[ 1- \frac{ 1- r_{\! \rm g}/r }{ 1- r_{\! \rm g}/r_0 }
\Biggr]^{1/2}.
\end{equation}

    Integration~(\ref{drdt}) yields the following expression for
the time $t-t_0$ of free fall from the point~$r_0$ (a particle at rest)
at the instant of time to a point with coordinate $r < r_0$:
\begin{eqnarray}
t - t_0 &=& \! \frac{r_{\! \rm g}}{c} \left\{ \sqrt{x_0-1 \mathstrut}
\left[ (2+x_0) \arctan \sqrt{ \frac{x_0 - x}{x}} +
\sqrt{x (x_0 -x) \mathstrut} \right]  \right.+
\nonumber  \\
&+& \!\!\! \left. 2 \ln \Biggl(\! \sqrt{(x_0 -1) \frac{x}{x_0}} +
\sqrt{1- \frac{x}{x_0}}\, \Biggr) - \ln |x-1| \right\},
\label{mt}
\end{eqnarray}
where $x_0 = r_0/r_{\! \rm g}$, and\, $x = r/r_{\! \rm g}$.
    The free-fall time obviously increases logarithmically in~$r-r_{\! \rm g}$
with no limits for $ x \to 1 $,\, i.e.\, $ r \to r_{\! \rm g} $.

    It might be possible to assume that, during this infinite Schwarzschild
time, the light rays from events that are arbitrary remote in space and time
could catch up with the freely falling astronaut.
    Let us make sure, however, that this is not the case.
    It should be noted, first of all, that the proper time of the astronaut
falling upon a black hole is finite.
    Indeed, for the proper time~$\tau - \tau_0 $ of motion from~$r_0$
to the point with radial coordinate~$r$ we obtain from Eqs.~(\ref{SG}):
    \begin{equation} \label{mts}
\tau - \tau_0 = \frac{1}{c} \int_r^{r_0}
\frac{d r}{\sqrt{\varepsilon^2 -1 + r_{\! \rm g}/r}} \,.
\end{equation}
    If the free fall occurs from the state at rest, then one has
    \begin{equation} \label{mtsX}
\tau - \tau_0 = \frac{r_0}{c} \, \sqrt{\frac{r_0}{r_{\! \rm g}}}
\left( \arctan \sqrt{ \frac{r_0}{r} -1 }
+ \sqrt{ \frac{r}{r_0} - \frac{r^2}{r_0^2} }\, \right).
\end{equation}
    Notice that time interval~(\ref{mtsX}) is exactly the same as
the appropriate free-fall time in Newton gravitational theory!

    Now consider the radial motion of a light ray.
    From the condition~$ds =0$, we get
\begin{equation} \label{rtf}
\frac{dr}{c\, dt} = \pm \left( 1- \frac{r_{\! \rm g}}{r} \right),
\end{equation}
which implies the photon propagation time from~$r_0$ to~$r$:
    \begin{equation} \label{gt}
t - t_s = \frac{r_0 - r}{c} +
\frac{r_{\! \rm g}}{c} \ln \left| \frac{r_0 -r_{\! \rm g}}{r-r_{\! \rm g}}
\right|,
\end{equation}
where $ t_s$ is the time of the photon start.
    Thus, the photon propagation time in the Schwarzschild coordinates
increases logarithmically in $r-r_{\! \rm g}$ as $ r \to  r_{\! \rm g}$.

\begin{figure}[ht]
\centering
   \includegraphics[width=10cm]{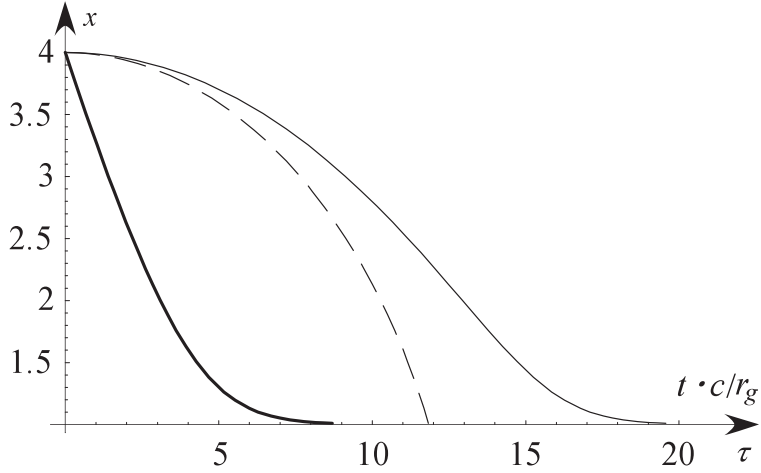}\\
\caption{\ The dependence $ x(t) $ and $ x(\tau) $ for a massive particle
falling upon a black hole and a photon (the thick solid curve).}
  \label{SchFig}
\end{figure}

    Figure~\ref{SchFig} plots the coordinate~$x$ (in units of
the Schwarzschild radius) as a function of time~$t$\, for a massive particle
and a photon (the thin and thick solid lines, respectively)
starting their motion at the point with~$r_0= 4 r_{\! \rm g}$.\
The dependence of the coordinate on the proper time~$\tau$\, of the massive
particle is shown by the dashed line.

    Subtracting expression~(\ref{gt}) from formula~(\ref{mt})
gives the answer to the following question:
At which instant of time~$t_s$ should a light signal be sent from
point~$r_0$\, in the radial direction to catch up with the freely falling
`observer' at a value of the Schwarzschild radius~$r<r_0$, who started his
motion with zero initial velocity from point~$r_0$\, at some instant of
time~$t_0 < t_s$?
    The answer follows as
\begin{eqnarray}
t_s - t_0 &=& \frac{r_{\! \rm g}}{c} \left[ (2+x_0)\sqrt{x_0-1 \mathstrut}\,
\arctan \sqrt{ \frac{x_0 - x}{x}} \right. +
\label{eq1}  \\ \nonumber
    &+& \left. \sqrt{x_0-x \mathstrut} \left( \sqrt{(x_0 - 1)\, x} -
\sqrt{x_0-x \mathstrut} \right)
+2 \ln \left( \sqrt{\frac{x}{x_0}} +
\sqrt{\frac{x_0 - x}{(x_0 - 1)\, x_0}}\, \right)  \right].
 \end{eqnarray}

    Proceeding in expression~(\ref{eq1}) to the limit $x \to 1$,\,
i.e. $r \to r_{\! \rm g} $,
we find how late the light can be emitted from the starting point of
the freely falling massive observer to be detected before the observer
crosses the horizon:
\begin{equation} \label{eq2}
t_s - t_0 =\frac{r_{\! \rm g}}{c} \left[ (2+x_0)\sqrt{x_0-1 \mathstrut}\,
 \arctan  \sqrt{x_0 - 1 \mathstrut}
+2 \ln 2 - \ln x_0 \right].
\end{equation}

    Thus, the limit is finite and before crossing the black hole horizon
there is no possibility of seeing the infinite future events occurring near
the starting point of the free fall.

    In Newtonian theory, the corresponding expression for~$t_s - t_0$\,
takes the form
    \begin{equation} \label{eq1n}
t_s - t_0 = \frac{r_{\! \rm g}}{c} \left[ x_0^{3/2} \arctan \sqrt{\frac{x_0-x}{x}}
+ \sqrt{x_0 x (x_0-x)} -(x_0-x) \right].
\end{equation}
    At large values of~$x_0/x= r_0/r \gg 1$,\,
both formulas~(\ref{eq1}) and~(\ref{eq1n}) give the same result
    \begin{equation} \label{eq4}
t_s - t_0 \sim \frac{\pi}{2}\, \frac{r_0}{c}\,
\sqrt{ \frac{\mathstrut r_0}{r_{\! \rm g}} } \,.
\end{equation}

    Let us consider another possible case of an astronaut falling upon
a black hole and seeing the future of the Universe.
    Instead of freely falling upon the black hole, the astronaut `lingers'
in some close orbit and rotates about it~\cite{Rees}.
    Here, the situation can be similar to the twins paradox:
in a short time interval, the astronaut will be able to observe processes
occurring over a rather long period of time in the vicinity of the Earth.
    But this is not the infinite future of the Universe!
    In addition, note that circular orbits with a radius smaller
than $r=3 r_{\! \rm g} $\, cease to be stable~\cite{ZeldovichNovikovTTES}.
    The velocity of travel in the last marginally stable circular orbit
equals~$c/2$,\, and hence the Lorentz time dilation here is insignificant.

    Let us go back to the question of the infinite time the astronaut needs
to approach the black hole horizon from the point of view of
the Schwarzschild's observer on the Earth and the astronaut's finite proper
time before crossing the horizon.
    Does that mean the `relativity of history'?
    Could it be that there is no unique history of the astronaut's motion
towards the black hole?
    If one understands history as a world line, it is clear that it is
unique and has a finite length.
    Another `history' possesses infinite length --- the world line of clocks
of the Schwarzschild's observer.
    So there is no relativity of history!

    Thus, we have considered the simplest case of a nonrotating, noncharged
black hole and the motion of an observer up to the horizon only.
    But, perhaps, when falling under the horizon towards the singularity
(which, as is well known, cannot be observed from outside the horizon)
all the future history of the world can become available to the brave
astronaut who left this world forever?
    Or maybe one can see the future in the case of charged or rotating
black holes?

    To explore these possibilities, a more complicated analysis of the global
structure of solutions describing black holes is needed.
    The interested reader should familiarize himself with
books~\cite{NovikovFrolov, Chandrasekhar, MTW, HawkingEllis} for more detail.
    Here we just want to bring up some basic facts which are relevant to
the problems considered.

\vspace{11pt}
\section{A fall under the horizon}

    Formula~(\ref{Sch}) becomes senseless when the infalling observer
crosses the horizon at~$r=r_{\! \rm g}$, which is related to the inadequacy
of the coordinate system.
    However, there are Kruskal--Szekeres coordinates which allow one to write
out the Schwarzschild solution both outside and inside the black hole horizon.

\begin{figure}[ht]
\centering
   \includegraphics[width=10cm]{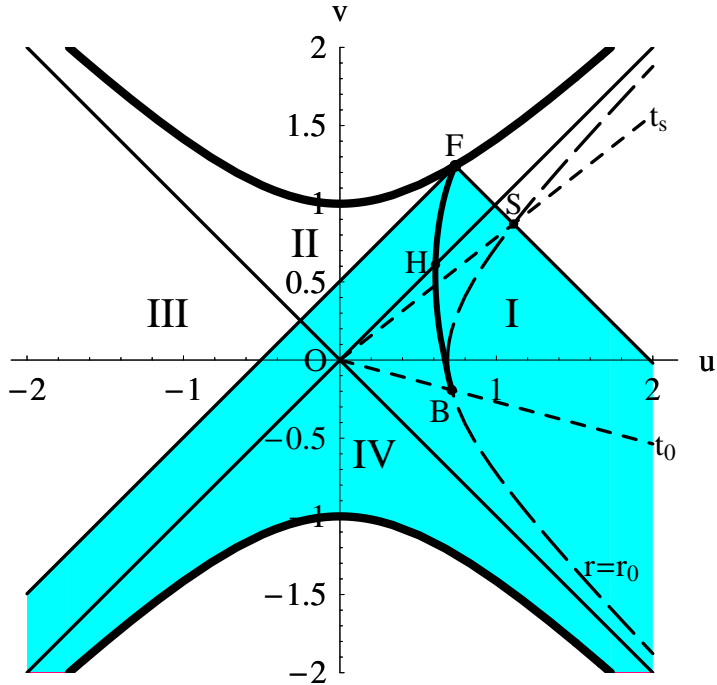}\\
\caption{\ Free fall upon a black hole in the Kruskal--Szekeres coordinates.}
  \label{Krus}
\end{figure}

    Let us introduce new coordinates $u,\ v $,\, such that
    \begin{equation} \label{UVxt}
u = \sqrt{\mathstrut x-1}\, \exp \! \left( \frac{x}{2} \right)
\cosh \frac{c t}{2 r_{\! \rm g}}\ , \ \ \ \ \ \
v = \sqrt{\mathstrut x-1}\, \exp \! \left( \frac{x}{2} \right)
\sinh \frac{c t}{2 r_{\! \rm g}}\ ,
\end{equation}
where $x \equiv r/r_{\! \rm g} >1 $.\,
    Here, clearly, the inequalities\ $u>|v| \ge 0$\, are valid.
    In these coordinates, the Schwarzschild metric~(\ref{Sch}) takes the form
    \begin{equation} \label{dsKS}
ds^2 = r_{\! \rm g}^2 \, \biggl[\, \frac{4}{x \exp x}
\left( d v^2 -d u^2 \right) - x^2 \left( d \vartheta^2 +
\sin^2 \vartheta \, d \varphi^2 \right) \biggr].
\end{equation}
    Next, let us assume that the coordinates $u$, $v$\, are changing
from~$-\infty$ to~$+\infty$,\, and $x>0$\, is a function of $u$, $v$\,
given by the following equation
    \begin{equation} \label{u2v2}
u^2 - v^2 = (x-1) \exp x \,.
\end{equation}
    This is the Kruskal--Szekeres coordinate system.
    In these coordinates, the world lines can be conveniently depicted
in spacetime both inside and outside the black hole horizon
(Fig.~\ref{Krus}).

    Here, one considers an `eternal' black hole which actually has two
singularities (their equation is $v^2-u^2=1$;\,
they are shown by the upper and lower heavy hyperbolas in Fig.~\ref{Krus}),
hidden from the external observer under the horizon.
    The second (bottom) singularity is absent for black holes that originated
from stellar collapses.

    Mapping~(\ref{UVxt}) covers only one-fourth of the $(u,v)$-plane:
region {\bf I} in Fig.~\ref{Krus}.
    In region~{\bf II}, where $v>|u|$ and $0<x<1$, we define
    \begin{equation} \label{ScwII}
r\equiv r_{\! \rm g} x\,,\ \ \ \ \
t\equiv 2 \frac{r_{\! \rm g}}{c}\, \artanh \frac{u}{v}\,.
\end{equation}
    The inverse transformation takes the form
    \begin{equation} \label{UVxtII}
u = \sqrt{\mathstrut 1 - x}\, \exp \! \left( \frac{x}{2} \right)
\sinh \frac{c t}{2 r_{\! \rm g}}\ , \ \ \ \ \ \
v = \sqrt{\mathstrut 1 - x}\, \exp \! \left( \frac{x}{2} \right)
\cosh \frac{c t}{2 r_{\! \rm g}}\ .
\end{equation}

    In region~{\bf II}, one has\,
$0<r<r_{\! \rm g}$,\ $t \in (-\infty, +\infty)$,
and Kruskal--Szekeres metric~(\ref{dsKS}) takes the form of
the Schwarzschild metric~(\ref{Sch}).
    However, now (inside the horizon) the coordinate~$r$\,
becomes timelike, and~$t$\, becomes spacelike!
    Therefore, by denoting
$\eta=r/c$,\ $l=ct$,\, where\
$\eta \in (0,\, r_{\! \rm g} / c), \ \, l \in {\rm \mathbf{R}}$,
we write out the metric inside the horizon in the form
   \begin{equation} \label{ScwIInl}
d s^2 = \frac{c^2 d \eta^2}{r_{\! \rm g}/(c \eta) - 1}
- \left( \frac{r_{\! \rm g}}{c \eta}-1 \! \right) d l^2 -
(c \eta )^2 \Bigl( d \vartheta^2 +\sin^2 \vartheta \, d \varphi^2 \Bigr)\,.
\end{equation}
    The spacetime described by metric~(\ref{ScwIInl}) is quite unusual.
    The space of this `universe', i.e., the surface $\eta={\rm const} $,
is spherically symmetric but anisotropic.
    The direction along the $l$-axis is preferential.
    Surfaces $(\eta={\rm const}$, $l={\rm const})$
represent ${\rm \mathbf{S}}^2$ spheres.
    However, the coordinate~$l$ is not radial.
    It takes all real values and the metric~(\ref{ScwIInl}) does not depend
on it.
    The topology of spatial cross section $\eta={\rm const} $\,
is the~$ {\rm \mathbf{R}}^1 \times {\rm \mathbf{S}}^2$ topology.
    From outside the black hole it appears that the space inside
the black hole has a finite volume, but inside the black hole it turns out
that there is a world line of infinite length (a cylinder of infinite length).
    It is exactly in connection with this remarkable property that we must
agree with Ne'eman's note about `different realities'.
    The reality inside the black hole cannot be imagined from the point of
view of the reality outside it, although it can be understand!

    The radius of the sphere ${\rm \mathbf{S}}^2$ (it is equal to $c\eta$)
decreases with time and vanishes at~$\eta=0$,\, which corresponds to
the Schwarzschild singularity.
    This singularity is not a space point inside the black hole but
represents the disruption of time for all world lines inside it.
    The Schwarzschild singularity is spacelike, and for any observer under
the horizon it is located in the {\it future.}
    It is impossible for an observer to `see' the singularity inside
the black hole before his own catastrophic destruction.
    The opposite statement in Refs.~\cite{CherCHDV, AstronomiyaVekXXI}
is erroneous.

    In the Kruskal--Szekeres coordinates, radial geodesics along which
the light propagates are represented by straight lines inclined by~$45^\circ$
to the coordinate axes.
    So, the radial light cone in the Kruskal--Szekeres coordinates has
the same form as in special relativity.
    This property allows one to easily establish the causal link between
events using graphical representation in the Kruskal--Szekeres
coordinates~\cite{MTW}.
    Let us take advantage of this property to answer the question of which
light signals catch up with the infalling observer under the event horizon.
    The world line of the observer who started the free fall from
point~$r_0$\, at instant of time~$t_0$\, is shown by the line $\mathbf{BHF}$
in Fig.~\ref{Krus}.
    The point~$\mathbf{F}$ corresponds to the world line disruption at
the singularity.
    The causal past of the event~$\mathbf{F}$ is shown in green.
    In the Kruskal--Szekeres coordinates, the surfaces of constant
radius~$r$\, are shown by hyperbolas with asymptotes $ u = \pm v $,
and surfaces of constant time~$t$ are represented by straight lines passing
through the origin of the coordinates.
    Therefore, as indicated in Fig.~\ref{Krus}, until the tragic ruin
of the free-falling observer at the singularity at the instant~$\mathbf{F}$,
light rays emitted from the point~$\mathbf{B}$ of a free-fall origin no later
than the time~$t_s$\, corresponding to the line $\mathbf{OS}$ can catch up
with observer.
    Therefore, during the free fall inside the black hole up to
the singularity the observer cannot see the infinite future!


    It should be noted, however, that in a rotating black hole described
by the Kerr metric or in a charged black hole
(Reissner--Nordstr\"{o}m metric and Kerr--Newman metric), a phenomenon
formally shows up that could be described as the possibility of an observer
seeing all the future of the Universe external to the black hole.
    In addition to the event horizon, as in the Schwarzschild metric, here
a new horizon, the Cauchy horizon, appears.
    The Cauchy horizon inside a black hole is the boundary for prediction
of the evolution of physical fields from initial data in the external Universe.
    The future of an astronaut crossing such a horizon is unpredictable from
their past.
    In the Kerr metric, the Cauchy horizon represents a null (lightlike)
surface.
    The astronaut can approach this horizon after crossing the first one
(the event horizon).
    So, as shown in several textbooks
(see, for example, Ref.~\cite{Chandrasekhar}, \S~38), at the instant of
crossing the Cauchy horizon surface, ``...the person will witness,
in a flash, a panorama of the entire history of the external world,
in infinitely blue shifted rays.''
    Nevertheless, as started in Ref.~\cite{NovikovFrolov}, \S~12.2,
the infinite blue shift means such a large energy concentration that
``would lead to the reconstruction of the spacetime and to the emergence
of the true singularity of the spacetime.''

    Does that mean that inside such a black hole it is impossible to see
the infinite future of the external Universe?
    So far, there is no definite answer to this question.
    To analyze this problem, it is insufficient to consider the Kerr solution
only.
    An analysis of the evolution of the singularity under the action of
radiation falling into the black hole is in order.
    For example, in paper~\cite{PoissonIsrael90, Ori92} and in the new English
edition of V~P~Frolov and I~D~Novikov's book~\cite{NovikovFrolov},
it is shown that if one takes into account gravitational perturbations to
the Reissner--Nordstr\"{o}m or Kerr metric, the Cauchy horizon surface
becomes singular --- a new singularity emerges, which is different from
both the spacelike and timelike singularities in respective Schwarzschild
and Kerr metrics.
    This singularity belongs to the class of `weak'
singularities~\cite{Tipler77}.
    It was argued in paper~\cite{Ori92} that ``the tidal deformation
associated with the singularity is so small that it cannot damage the object,
and, in some conditions, it cannot even be detected before the curvature
becomes infinite.
    This reopens the question of whether a journey through the Cauchy horizon
of black holes is possible.''
    However, if one takes into account the inverse effect of external fields,
for example, the massless scalar field in some model
problems~\cite{Burko03, HKhNovikov05}, the null singularity can, under certain
conditions, evolve into a strong spacelike singularity.
    The case is also possible where two singularities, the null and
strong spacelike ones, exist simultaneously.

    Even if this singularity remains the null one, the question of
whether the astronaut can cross it lacks clear answer.
    When a strong spacelike singularity is present, the astronaut will be
destroyed by tidal forces.
    All these results were obtained for several model problems that allow
simple mathematical solutions.
    What the astronaut sees inside a real rotating black hole, which
cannot be described by the Kerr metric any more, is unclear.

\vspace{7pt}
{\bf Acknowledgments.}
The authors thank S V Krasnikov for the fruitful discussion.
This work was partially supported by the Russian Federation Ministry
of education and Science, grant RNP.2.1.1.6826.

\newpage


\begin{thebibliography}{99}

\bibitem{Cherepashchuk03}
Cherepashchuk A M \ {\it Usp. Fiz. Nauk\/} {\bf 173} 345 (2003)
[{\it Phys. Usp.\/} {\bf 46} 335 (2003)]

\bibitem{LogunovMst}
Logunov A A, Mestvirishvili M A \ {\it Relyativistskaya Teoriya Gravitatsii\/}
(The Relativistic Theory of Gravitation) (Moscow: Nauka, 1989)
[Translated into English (Moscow: Mir Publ., 1989)]

\bibitem{Schwarzschild16}
Schwarzschild K \ {\it Sitz. Preuss. Akad. Wiss. Berlin\/} 189 (1916)

\bibitem{Kerr63}
Kerr R P \ {\it Phys. Rev. Lett.\/} {\bf 11} 237 (1963)

\bibitem{Reissner16}
Reissner H \ {\it Ann. der Phys.\/} {\bf 355} 106 (1916);\\
Nordstr\"{o}m G \ {\it Kon. Nederland. Akad. Wet. Proc.\/}
{\bf 20} 1238 (1918)

\bibitem{Newmanetal65}
Newman E T et al. \
{\it J. Math. Phys.} {\bf 6} 918 (1965)

\bibitem{NovikovFrolov}
Novikov I D, Frolov V P \ {\it Fizika Chernykh Dyr\/}
(Black Hole Physics) (Moscow: Nauka, 1986)
[Translated into English:
Frolov V P, Novikov I D \ {\it Black Hole Physics: Basic Concepts and
New Developments\/} (Dordrecht: Kluwer, 1998)]

\bibitem{LL_II}
Landau L D, Lifshitz E M \ {\it Teoriya Polya\/}
(The Classical Theory of Fields) (Moscow: Nauka, 1988)
[Translated into English: (Oxford: Pergamon Press, 1975)]

\bibitem{Neeman}
Ne'eman Yu \ {\it Found. Phys. Lett.\/} {\bf 7} 483 (1994)

\bibitem{Kruskal60}
Kruskal M D \ {\it Phys. Rev.\/} {\bf 119} 1743 (1960);\\
Szekeres G \ {\it Publ. Mat. Debrecen\/} {\bf 7} 285 (1960)

\bibitem{Eizenshtedt82}
Eisenstaedt J\, {\it Arch. Hist. Exact Sci.\/} {\bf 27} 157 (1982)
[Translated into Russian, in {\it Ein\-shtei\-novskii Sbornik (1984---1985)\/}
(Exec. Ed.\, I Yu Kobzarev) (Moscow: Nauka, 1988) p.~148]    

\bibitem{CherCHDV}
Cherepashchuk A M \ {\it Chernye Dyry vo Vselennoi}
(Black Holes in the Universe) (Fryazino: Vek~2, 2005)

\bibitem{AstronomiyaVekXXI}
Cherepashchuk A M, in\, {\it Astronomiya: Vek XXI\/}
(Astronomy: the 21st Century) (Comp. Ed. V G Surdin)
(Fryazino: Vek~2, 2007) p~219

\bibitem{HawkingUnNut}
Hawking S \ {\it The Universe in a Nutshell\/}
(New York: Bantam Books, 2001)
[Translated into Russian (St.\,Petersburg: Amfora, 2008)]

\bibitem{Redze}
Regge T \ {\it Cronache Dell'Universo} (Torino: P. Boringhieri, 1981)
[Translated into Russian (Moscow: Mir Publ., 1985) p.~34]

\bibitem{Chandrasekhar}
Chandrasekhar S \ {\it The Mathematical Theory of Black Holes\/}
(Oxford: Oxford Univ. Press, 1983)
[Translated into Russian (Moscow: Mir Publ., 1986)]

\bibitem{Rees}
Rees M \ {\it Our Cosmic Habitat\/}
(Princeton: Princeton Univ. Press, 2001)
[Translated into Russian (Moscow--Izhevsk: Inst. Komp'yut. Issled., 2002) p.~96]

\bibitem{ZeldovichNovikovTTES}
Zel'dovich Ya B, Novikov I D \ {\it Teoriya Tyagoteniya i Evolyutsiya Zvezd\/}
(Gravitational Theory and Evolution of Stars) (Moscow: Nauka, 1971)

\bibitem{MTW}
Misner C W, Thorne K S, Wheeler J A \ {\it Gravitation\/}
(San Francisco: W.H. Freeman, 1973)
[Translated into Russian (Moscow: Mir, 1977)]

\bibitem{HawkingEllis}
Hawking S W, Ellis G F R \ {\it The Large Scale Structure of Space-Time\/}
(Cambridge: Cambridge Univ. Press, 1973)
[Translated into Russian (Moscow: Mir, 1977)]

\bibitem{PoissonIsrael90}
Poisson E, Israel W \ {\it Phys. Rev. D\/} {\bf 41} 1796 (1990)

\bibitem{Ori92}
Ori A \ {\it Phys. Rev. Lett.\/} {\bf 68} 2117 (1992)

\bibitem{Tipler77}
Tipler F J \ {\it Phys. Lett. A\/} {\bf 64} 8 (1977)

\bibitem{Burko03}
Burko L M \ {\it Phys. Rev. Lett.\/} {\bf 90} 121101 (2003)

\bibitem{HKhNovikov05}
Hansen J, Khokhlov A, Novikov I
\ {\it Phys. Rev. D\/} {\bf 71} 064013 (2005)

\end{thebibliography}
\end{document}